\documentclass{aa}  
\usepackage{graphicx,dblfloatfix}
\usepackage{txfonts}
\usepackage{float}

\usepackage{xcolor}

\bibpunct{(}{)}{;}{a}{}{,} 

\begin{document}

   \title{Viscous Heating and Boundary Layer Accretion in the Disk of Outbursting Star FU\,Orionis}


   \author{Aaron Labdon
          \inst{1}
          \and
          Stefan Kraus
          \inst{1}
          \and  
          Claire L.\ Davies
          \inst{1}
          \and
          Alexander Kreplin
          \inst{1}
          \and
          John D.\ Monnier
          \inst{2}
          \and
          Jean-Baptiste Le\ Bouquin
          \inst{3}
          \and
          Narsireddy\ Anugu
          \inst{1,2,4}
          \and
          Theo ten Brummelaar
          \inst{5}
          \and
          Benjamin Setterholm
          \inst{2}
          \and
          Tyler Gardner
          \inst{2}
          \and
          Jacob Ennis
          \inst{2}
          \and
          Cyprien Lanthermann
          \inst{6,3}
          \and
          Gail Schaefer
          \inst{5}
          \and
          Anna Laws
          \inst{1}
          }

   \institute{
   (1) University of Exeter, School of Physics and Astronomy, Astrophysics Group, Stocker Road, Exeter, EX4 4QL, UK\\
   (2) University of Michigan, Department of Astronomy, S University Ave, Ann Arbor, MI 48109, USA\\
   (3) Institut de Planetologie et d’Astronomie de Grenoble, Grenoble 38058, France\\
   (4) Steward Observatory, Department of Astronomy, University of Arizona, Tuscon, USA\\
   (5) The CHARA Array of Georgia State University, Mount Wilson Observatory, Mount Wilson, CA 91023, USA\\
   (6) Instituut voor Sterrenkunde, KU Leuven, Celestijnenlaan 200D, 3001 Leuven, Belgium 
   }


 
  \abstract
   {FU\,Orionis is the archetypal FUor star, a subclass of young stellar object (YSO) that undergo rapid brightening events, often gaining 4-6 magnitudes on timescales of days. This brightening is often associated with a massive increase in accretion; one of the most ubiquitous processes in astrophysics from planets and stars to super-massive black holes. We present multi-band interferometric observations of the FU\,Ori circumstellar environment, including the first J-band interferometric observations of a Y SO.
   }
   {We investigate the morphology and temperature gradient of the inner-most regions of the accretion disk around FU\,Orionis. We aim to characterise the heating mechanisms of the disk and comment on potential outburst triggering processes.
   }
   {Recent upgrades to the MIRC-X instrument at the CHARA array allowed the first dual-band J and H observations of YSOs. Using baselines up to 331\,m, we present high angular resolution data of a YSO covering the near-infrared bands J, H, and K. The unprecedented spectral range of the data allows us to apply temperature gradient models to the innermost regions of FU\,Ori.
   }
   {We spatially resolve the innermost astronomical unit of the disk and determine the exponent of the temperature gradient of the inner disk to $T\propto r^{-0.74\pm0.02}$. This agrees with theoretical work that predicts $T\propto r^{-0.75}$ for actively accreting, steady state disks, a value only obtainable through viscous heating within the disk. We find a disk which extends down to the stellar surface at $0.015\pm0.007\,\mathrm{au}$ where the temperature is found to be $5800\pm700\,\mathrm{K}$ indicating boundary layer accretion. We find a disk inclined at $32\pm4^\circ$ with a minor-axis position angle of $34\pm11^\circ$.
   }
   {We demonstrate that J-band interferometric observations of YSOs are feasible with the MIRC-X instrument at CHARA. The temperature gradient power-law derived for the inner disk is consistent with theoretical predictions for steady-state, optically thick, viciously heated accretion disks.
   }


   \maketitle
%

\section{Introduction} \label{sec:intro}

    Accretion onto astronomical objects is one of the most fundamental processes in astrophysics and facilitates mass transport onto a wide range of astrophysical objects, from planets and stars to super-massive black holes \citep{Lin96}. The mass transport proceeds through accretion disks, where viscosity converts angular momentum into thermal energy, thus enabling the mass infall \citep{Pringle72}. A key prediction is that the disk viscosity should actively heat the disk, where the radial temperature profile has been predicted as far back as the 1970’s \citep{Shakura73,Shibazaki75,Hartmann85}.
    
    The radial temperature gradient of circumstellar disks are determined by the heating mechanisms that power them. The two primary heating mechanisms in protoplanetary disks are the reprocessing of stellar radiation and viscous heating \citep{DAlessio05}. Stellar radiation is reprocessed through absorption, re-emission, and scattering of photons and is most effective in the outer layers of the disk owing to the optical depth of material\citep{Natta01}. Viscous heating on the other hand is thought to be confined to the mid-plane where it is thought to be driven by turbulence and instabilities in the disk material, such as magneto-rotation instabilities (MRI) \citep{Balbus98}. However, the presence and nature of viscosity is highly debated and relatively unconstrained by current observational data. 

    The appreciation of accretion and viscousity processes in young stellar objects (YSOs) is vital to the understanding of both star- and planet-formation mechanisms. Active accretion disks have been observed around a wide range of YSO classes, however YSOs are known to be 10-100 times less luminous than expected from steady-state accretion scenarios. Particularly given the accretion rates of the order $10^{-7}$ to $10^{-8}\,\mathrm{M_\odot yr^{-1}}$ observed around many YSOs. This raises the possibility that accretion is not consistent across the early states of stellar evolution, but is episodic \citep{Kenyon95,Evans09}. Such scenarios may manifest in outbursting events.


    
    Many stars are known to undergo episodic accretion events of which FU\,Orionis (FUor) stars are the best known. FUors are characterised by rapid brightening events \citep{Audard14} followed by a protracted period of dimming, on the order of decades to centuries to return to the quiescence state  \citep{Hartmann96,Herbig07,kra16}. In such an outburst the accretion rates can increase
    to $\sim 10^{-4}\,\mathrm{M_\odot yr^{-1}}$. It is now thought that most YSOs stars exhibit episodic accretion, undergoing at least one or more outburst events throughout their lifetime \citep{Hartmann96,Audard14.}
    
     \begin{table}[t!]
        \caption{\label{table:Stellar}Stellar parameters of the FU\,Orionis system. The mass and radii of the FU Ori N are derived in this work (see Section\,\ref{GeoMod}). }
        \centering
        \begin{tabular}{c c c} 
            \hline
            \noalign{\smallskip}
            Parameter  &  Value &   Reference   \\ [0.5ex]
            \hline
            \noalign{\smallskip}
            FU Ori North & & \\
            \hline
            \noalign{\smallskip}
            RA (J2000) &  $05~45~22.362$   & (1)  \\
            DEC (J2000) & $+09~04~12.31$  & (1)          \\
            Distance & $416\pm8~\mathrm{pc}$ & (1)  \\
            $H_{mag} $ & $5.68\pm0.1$ & (2) \\ 
            $A_V$ & 1.4 & (3) \\
            Mass & $0.6\pm0.1~\mathrm{R_\odot}$ & (4) \\[1ex]
            \hline
            \noalign{\smallskip}
            FU Ori South & & \\
            \hline
            \noalign{\smallskip}
            Separation & $0".5 (225\,\mathrm{au})$ & (2)\\ 
            Mass & $1.2\,\mathrm{M_\odot}$ & (2)\\
            $A_V$ & 8-12 & (3)\\  [1ex] 
            \hline
        \end{tabular}
        \tablebib{
        (1) \citet{BailerJones18}; (2)~\citet{Beck12}; (3)~\citet{Pueyo12}; (4)~\citet{Perez20} 
        }
    \end{table}
    
    The triggering mechanism of outbursts in FUor-type stars is not well understood. Various scenarios have been proposed to explain the massive increase in accretion rate seen in these objects. \citet{Vorobyov05,Vorobyov06} propose gravitational instabilities on large scales cause the disk to fragment and for clumps of material to fall onto the central star. \citet{Bell94} suggest that a thermal instability in the very inner regions ($<0.1$\,au) could be sufficient to cause outbursts of this magnitude. \citet{Bonnell92} propose that a binary companion on a highly eccentric orbit could perturb the disk and cause repeated outburst of accretion. Similarly, \citet{Reipurth04} suggest that FUor stars are newly created binary systems, where the two stars become bound following the breakup of larger multiple systems. Such a scenario leads to the ejection of companions and the rapid infall of material. However, given the limited number of known FUors there is little consensus on the FUor outburst triggering mechanism.
    
    FU\,Orionis (FU\,Ori) is the archetypal FUor object located in the Lambda-Orion star forming region at a distance of $416\pm8\,\mathrm{pc}$ \citep{BailerJones18}. The full stellar parameters are listed in Table\,\ref{table:Stellar}. FU\,Ori is known to be a binary system with an actively accreting companion located $0.5"$ to the South. Despite being \textasciitilde$4$\,mag fainter (in the V band), FU\,Ori~S is thought to be the more massive object ($1.2\,\mathrm{M_\odot}$) \citep{Beck12}. Meanwhile the mass of the Northern component has been estimated to $0.3\pm0.1\,\mathrm{M_\odot}$ based on modelling of the Spectral Energy Distribution \citet[SED,][]{Zhu07}.
    In 1937 FU\,Ori~N underwent a rapid brightening event whereby its brightness increased from $15.5^m\pm0.5$ to $9.7^m\pm0.1$ \citep{Clarke05} in the photographic system (similar spectral response to Johnson B filter). Since reaching a peak shortly after outburst its magnitude has decreased steadily at around $0.0125^{m}$ per year in B band ($540\,\mathrm{nm}$).
    
    FU\,Ori is one of the best-studied YSOs due to its high magnitude and unique nature. Scattered light observations by \citet{Takami14} and \citet{Laws20} with Subaru and GPI respectively reveal a large spiral arm structure in the north-west of the disk extending around 200\,au that has been attributed to a gravitational instability in the outer disk. Additionally, they reveal a stripe/shadow in the Northern disk and a diffuse outflow extending to the east, which is tentatively attributed to a jet in polar direction. \citet{Eisner11} observed FU\,Ori using the Keck interferometer in high resolution K-band mode ($\lambda = 2.15\,-\,2.36$, $R=2000$). They found a temperature gradient of $T\propto R^{-0.95}$ for the inner disk using a single baseline. Additionally, \citet{Liu19} presented medium resolution GRAVITY observations ($\lambda = 2.0\,-\,2.45$, $R=500$) which they model as an off-centre face-on Gaussian object. They also highlight $\mathrm{H_2O}$ and CO absorption features that are consistent with a $\Dot{M}\sim10^{-4}\,\mathrm{M_\odot yr^{-1}}$ accretion disk model. ALMA observations taken by \citet{Perez20} show a disk inclined at $\sim37^\circ$ at a position angle of $134\pm2^\circ$, both values are aligned to a disk detection around the southern component. Using methods outlined by \citet{Zhu07}, \citet{Perez20} were able to constrain the stellar parameters of the northern component based on the disk geometry. They find a stellar mass of $0.6\,M_\odot$, with a mass accretion rate of $3.8\times10^{-5}\,\Dot{M}yr^{-1}$. They go further to derive an inner disk radius of $3.5\,R_\odot$ or $0.016\,\mathrm{au}$. Mid-infrared (MIR) interferometry in the N-band and SED analysis by \citet{Quanz06} derive a temperature gradient for the outer disk ($>3\,\mathrm{au}$) which is in good agreement to what can be found for isothermal flared disks in theoretical work \citep{Kenyon87}.
    
     \renewcommand{\thefigure}{1}
    \begin{figure}[b]
        \centering
        \includegraphics[scale=0.35]{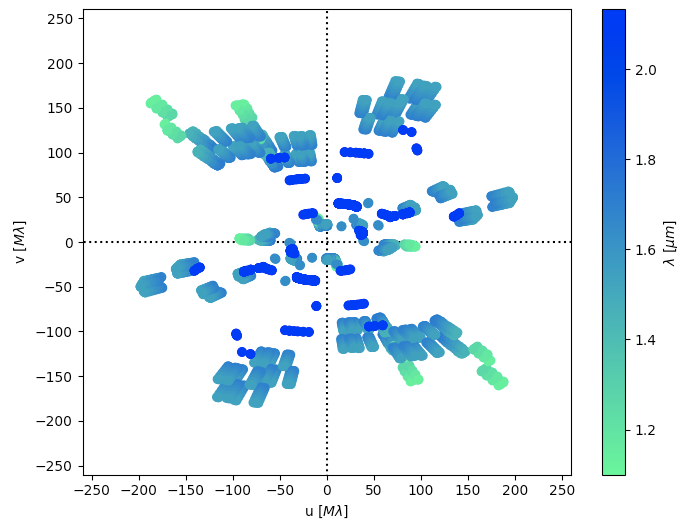}
        \caption{The uv-plane of all observations detailed in Table\,\ref{table:LOG}. The H+J dual-band observations were obtained with CHARA/MIRC-X. The H-band only observations were obtained with CHARA/MIRC-X with $R=50$ spectral resolution and with VLTI/PIONIER in free mode, while the K band observations were obtained with CHARA/CLIMB and PTI.}
        \label{fig:uv_map}
    \end{figure}
    
     \begin{table*}[ht]
    \caption{\label{table:LOG}Observing log of all instruments, data spanning 10 years from 1998 to 2019. Calibrator identifiers and uniform disk diameters (UDDs) are shown in Appendix\,\ref{AppA}.}
    \centering
    \begin{tabular}{c c c c c c} 
        \hline
        \noalign{\smallskip}
        Date  &  Beam Combiner & Filter &   Stations  & Pointings & Calibrator (see Table\,\ref{table:Cals}) \\ [0.5ex]
        \hline
        \noalign{\smallskip}
        2019-11-07 &  CHARA/MIRC-X  & J+H band & S1-S2-W1-W2 & 2 & (1) \\
        \hline
        \noalign{\smallskip}
        2018-11-27 &  CHARA/MIRC-X & H band  & S1-S2-E1-E2-W1-W2   & 2 & (2), (3) \\
        2019-11-06 &  CHARA/MIRC-X  & H band & S1-S2-E1-W1-W2    & 1 & (4) \\
        \hline
        \noalign{\smallskip}
        2017-12-25 &  VLTI/PIONIER  & H band & D0-G2-J3-K0   & 1 & (5), (6) \\
        \hline
        \noalign{\smallskip}
        2016-11-24 & VLTI/GRAVITY & K band & D0-G2-J3-K0 & 2 & (7), (8) \\
        2016-11-25 & VLTI/GRAVITY & K band & D0-G2-J3-K0 & 2 & (8) \\
        \hline
        \noalign{\smallskip}
        2010-10-02 &  CHARA/CLIMB  & K band & S2-E2-W2 & 3 & (3) ,(11) \\
        2010-11-30 &  CHARA/CLIMB  & K band & S1-E1-W1 & 2 & (3)  \\
        2010-12-04 &  CHARA/CLIMB  & K band & S2-E1-W2 & 5 & (10)\\
        2011-10-27 &  CHARA/CLIMB  & K band & S2-E2-W2 & 1 & (3)  \\
        2011-10-29 &  CHARA/CLIMB  & K band & S1-E1-W1 & 1 & (3)  \\[1ex] 
        \hline
        \noalign{\smallskip}
        1998-11-14 &  PTI  & K band & N-S   & 2 & (13) , (12),(13), (14) \\
        1998-11-16 &  PTI  & K band & N-S   & 2 & (13), (12), (14) \\
        1998-11-17 &  PTI  & K band & N-S   & 12 & (13), (12), (13) \\
        1998-11-19 &  PTI  & K band & N-S   & 5 & (11), (12), (15), (13) \\
        1998-11-22 &  PTI  & K band & N-S   & 9 & (11), (12), (13), (15) \\
        1998-11-23 &  PTI  & K band & N-S   & 4 & (11), (12), (13) \\
        1998-11-24 &  PTI  & K band & N-S   & 6 & (11), (12),(13) \\
        1998-11-25 &  PTI  & K band & N-S   & 6 & (11), (12), (13) \\
        1998-11-26 &  PTI  & K band & N-S   & 5 & (11), (13) \\
        1998-11-27 &  PTI  & K band & N-S   & 7 & (11), (13) \\
        1999-11-24 &  PTI  & K band & N-S   & 11 & (11), (13) \\
        1999-11-25 &  PTI  & K band & N-S   & 1 & (11), (15), (16), (17) \\
        1999-11-26 &  PTI  & K band & N-S   & 6 & (11), (13), (16), (17) \\
        1999-11-27 &  PTI  & K band & N-S   & 16 & (11), (13), (15), (16), (17) \\
        1999-11-28 &  PTI  & K band & N-S   & 21 & (11), (13), (15), (16), (17) \\
        2000-11-19 &  PTI  & K band & N-S   & 9 & (11), (17), (13), (18) \\
        2000-11-20 &  PTI  & K band & N-W   & 14 & (11), (17), (13) \\
        2000-11-22 &  PTI  & K band & N-W   & 3 & (11), (17),(13) \\
        2000-11-26 &  PTI  & K band & N-W   & 13 & (11), (17), (19), (13) \\
        2003-11-19 &  PTI  & K band & N-W   & 3 & (19), (11),(18) \\
        2003-11-20 &  PTI  & K band & N-S   & 2 & (19), (11), (18) \\
        2003-11-21 &  PTI  & K band & S-W   & 2 & (19), (11), (18) \\
        2003-11-27 &  PTI  & K band & S-W   & 1 & (19), (11), (18) \\
        2004-10-12 &  PTI  & K band & N-W   & 3 & (19), (11), (20), (18) \\
        2004-11-12 &  PTI  & K band & S-W   & 6 & (11), (19), (20), (18) \\
        2004-12-12 &  PTI  & K band & N-W   & 6 & (11), (19), (20), (18) \\
        2008-10-25 &  PTI  & K band & N-W   & 5 & (11), (13), (17) \\
        2008-11-14 &  PTI  & K band & N-S   & 1 & (11), (13), (17), (15) \\
        \hline
    \end{tabular}
    \end{table*}
    
    In non-outbursting low-mass YSO (T\,Tauri stars) the mass-infall is believed to proceed through magnetospheric accretion columns operating between the inner disk and the photosphere \citep{Bouvier07}.  It is not yet understood how the accretion geometry differs in highly-accreting FUors, which also limits our understanding of the outburst-triggering mechanisms. In order to further our understanding of these processes we have conducted the first tri-waveband NIR interferometric study of a YSO at the CHARA array using MIRC-X in the J and H bands. These include simultaneous J- and H-band observations using MIRC-X, the first of their kind. This data is complemented with observations in the K-band from the "CLassic Interferometry with Multiple Baselines" (CLIMB) beam combiner. The observations are introduced in Section\,\ref{Observations}. Our modelling techniques and temperature gradient analysis are shown in Section\,\ref{GeoMod}. The details of our companion search are outlined in Section\,\ref{Comp} and we discuss the implications and draw conclusions in Sections\,\ref{discussion} and \ref{Conclusion}, respectively.

    \section{Observations} \label{Observations}
    
    To collect the data for this multi-wavelength study, a variety of instruments operating at a range of wavelengths were employed. A summary of the observations is provided in Table\,\ref{table:LOG} and the resultant uv-coverage is shown in Figure\,\ref{fig:uv_map}.

    \subsection{CHARA/MIRC-X H-band Observations}
    
     The primary instrument used in this study is the "Michigan InfraRed Combiner-eXeter" (MIRC-X), which is a six-telescope beam combiner instrument located at the CHARA array.
     The CHARA array is a Y-shaped interferometric facility that comprises six $1\,$m telescopes. It is located at the Mount Wilson Observatory, California, and offers operational baselines between $34$ and $331\,$m \citep{Brummelaar05}. The MIRC-X instrument \citep{Monnier04,Kraus18,Anugu20} is recording data routinely in H-band ($\lambda=1.63\,\mathrm{\mu m}, \Delta\lambda=0.35\,\mathrm{\mu m}$ with $R=\lambda/\Delta\lambda=50$) since late 2017 \citep[first results in][]{Kraus20}.
     
     Overall we obtained 4 independent H-band pointings of FU\,Ori with MIRC-X, using a mixture of 5 and 6-telescope configurations in bracketed calibrator-science concatenation sequences. A maximum physical baseline of $331\,$m was used, corresponding to a maximum resolution of $\lambda/(2B) = 0.50\,\mathrm{mas}$ [milliarcseconds], where $\lambda$ is the observing wavelength and $B$ is the projected baseline.
     
     The data were reduced using the MIRC-X data reduction pipeline v1.2.0\footnote{https://gitlab.chara.gsu.edu/lebouquj/mircx\_pipeline} to produce calibrated squared visibilites and closure phases. The MIRC-X detector is susceptible to bias in the bispectrum estimation as pointed out by \citet{Basden04}. The pipeline corrects for this with a method similar to Appendix C of that paper. The measured visibilities and closure phases were calibrated using interferometric calibrator stars observed alongside the target; the calibrators are listed in Table\,\ref{table:LOG} and their diameters are given in Appendix\,\ref{AppA}. The adopted uniform diameters (UDs) were obtained from JMMC SearchCal \citep{Bonneau06, Bonneau11}.
    
    \subsection{CHARA/MIRC-X J+H band Observations}
    
    Recent developments have allowed the wavelength of MIRC-X to be extended to allow for simultaneous H and J-band observations through the implementation of longitudinal dispersion compensators (LDCs) to correct for atmospheric dispersion between the two bands \citep{Berger03}. LDCs consist of a wedge of SF-10 glass which is moved across the beam to increase or decrease the thickness of glass depending on the total airpath of the interferometric system. At the time, the LDCs were tracked manually resulting in sub-optimal fringe contrast in the J band, however a consistent approach was used across calibrator and science stars to ensure accurate calibration could be obtained. Such observations were conducted for the first time in November 2019, with automated LDC control now in the late stage of development \citep{Anugu20}.
    
    The data presented in this paper represents the first successful J-band interferometric observations of a YSO. These dual-band observations correspond to 14 spectral channels across wavelengths $1.08\,\mathrm{to}\,1.27\,\mathrm{\mu m}$ and $1.41\,\mathrm{to}\,1.73\,\mathrm{\mu m}$. The gap in the band pass is due to the presence of the CHARA metrology laser at around $1.3\,\mathrm{\mu m}$, this was removed using a narrow-band 'notch' filter. 
    
    We obtained two J+H band pointings on FU\,Ori with MIRC-X in 2019, using a 4-telescope configuration in CAL-SCI concatenation sequences. Only a 4-telescope configuration corresponding to the lowest spatial frequency fringes can be used for dual-band observations. Recording with 5 or 6 telescopes would result in the highest spatial frequency being undersampled (sub-Nyquist) on the detector. Of the available baselines, a maximum physical baseline of $280\,$m was used corresponding to a maximum resolution of $0.34\,\mathrm{mas}$.
    
    These data were reduced using an adapted version of the MIRC-X data reduction pipeline v1.2.0. The UDs of the calibrator stars were obtained from JMMC SearchCal \citep{Bonneau06, Bonneau11}. The spectral dependence of the UDs between the J and H bands is small enough to be considered negligible by the calibration pipeline.

    \subsection{VLT/PIONIER H-band Observations}

    FU\,Ori was recorded with the PIONIER instrument \citet{PIONIER11}. PIONIER is a four telescope beam combiner operating in the H-band ($\lambda = 1.64\,\mathrm{\mu m}$) at the VLTI. Data was obtained in December 2017 without a dispersive element (FREE mode) and reduced using the standard PNDRS pipeline \citep{JBLB2011}.

    \subsection{VLTI/CLIMB K-band Observations}
    
    We present observations obtained with the CLIMB instrument \citep{Brummelaar13}, also located at the CHARA array. CLIMB is a three telescope beam combiner that was used to obtain near-infrared K-band data ($\lambda=2.13\,\mathrm{\mu m}, \Delta\lambda=0.35\,\mathrm{\mu m}$) between November 2009 and October 2011. The CLIMB data were reduced using pipelines developed at the University of Michigan that are optimised for recovering faint fringes from low visibility data.
    
    Archival K-band data were also available from the Palomar Testbed Interferometer \citep[PTI,][]{Colavita99} from 1998 to 2008 using a two-telescope beam combiner on 3 different physical baselines between $86$ and $110$\,m. The data on FU\,Ori was published in \citet{Malbet98,Malbet05}. These additional measurements complement the CHARA observations in the short to intermediate baseline range; the full uv coverage is shown in Figure\,\ref{fig:uv_map}.

    \subsection{Photometric Observations}
    
    Photometric observations were collected from a variety of sources in order to build up the spectral energy distribution (SED) of FU\,Ori. Where possible, care was taken to minimize the time difference between observations and the number of instruments used. In particular, photometric data taken during the 1998 to 2019 period of the interferometric observations was considered whenever possible. A full list of the photometric observations used can be seen in Appendix\,\ref{AppB}.

     \renewcommand{\thefigure}{2}
    \begin{figure}[ht]
        \centering
        \includegraphics[scale=0.4]{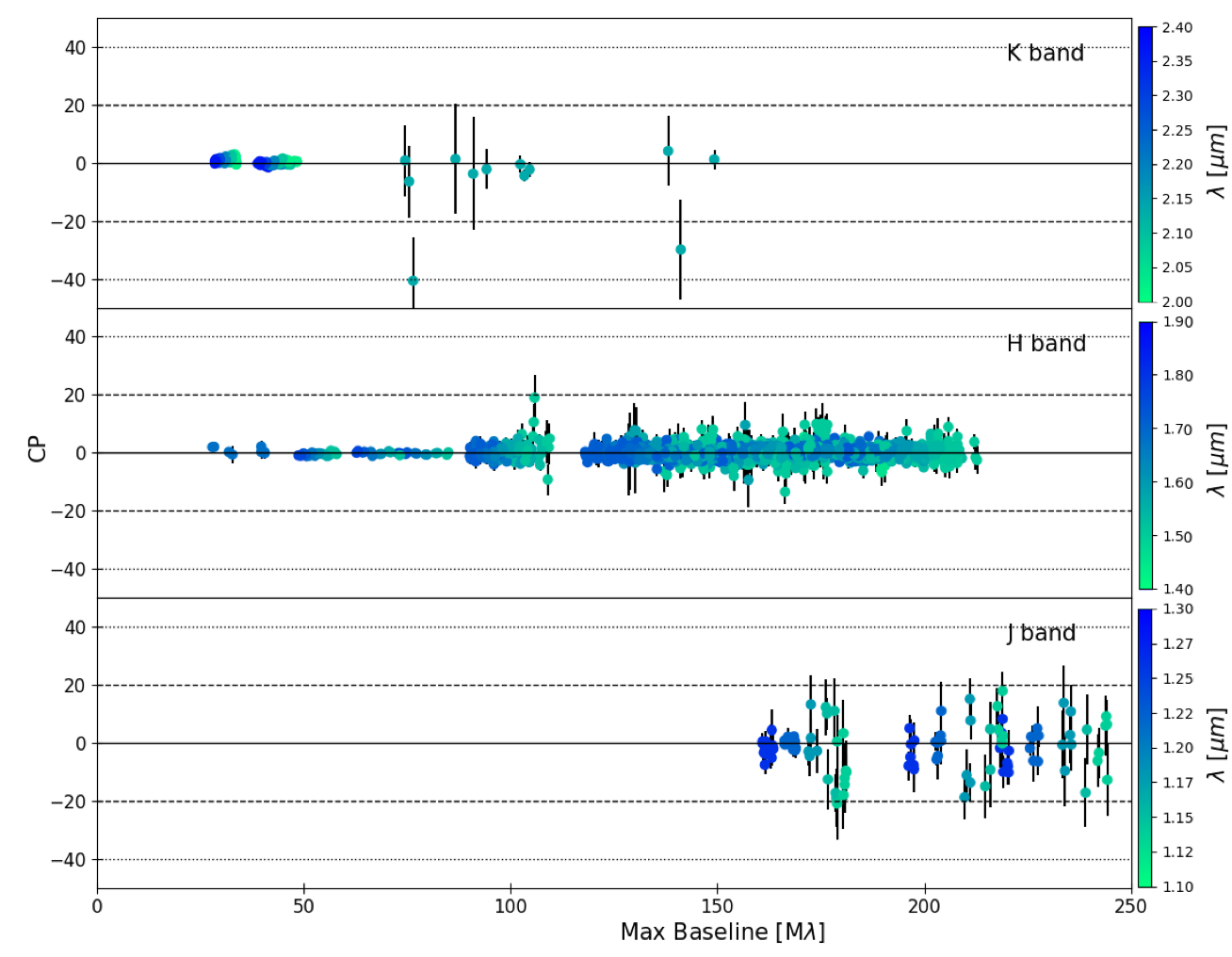}
        \caption{Closure phases for each waveband plotted against maximum baseline length, coloured according to observing wavelength. Shorter wavelengths in each band correspond to lighter green colours while longer wavelengths in each band are represented as dark blue.}
        \label{fig:CPs}
    \end{figure}

\section{Modelling and Results}\label{Modelling}
    
    \renewcommand{\thefigure}{3}
    \begin{figure*}
        \centering
        \includegraphics[scale=0.55]{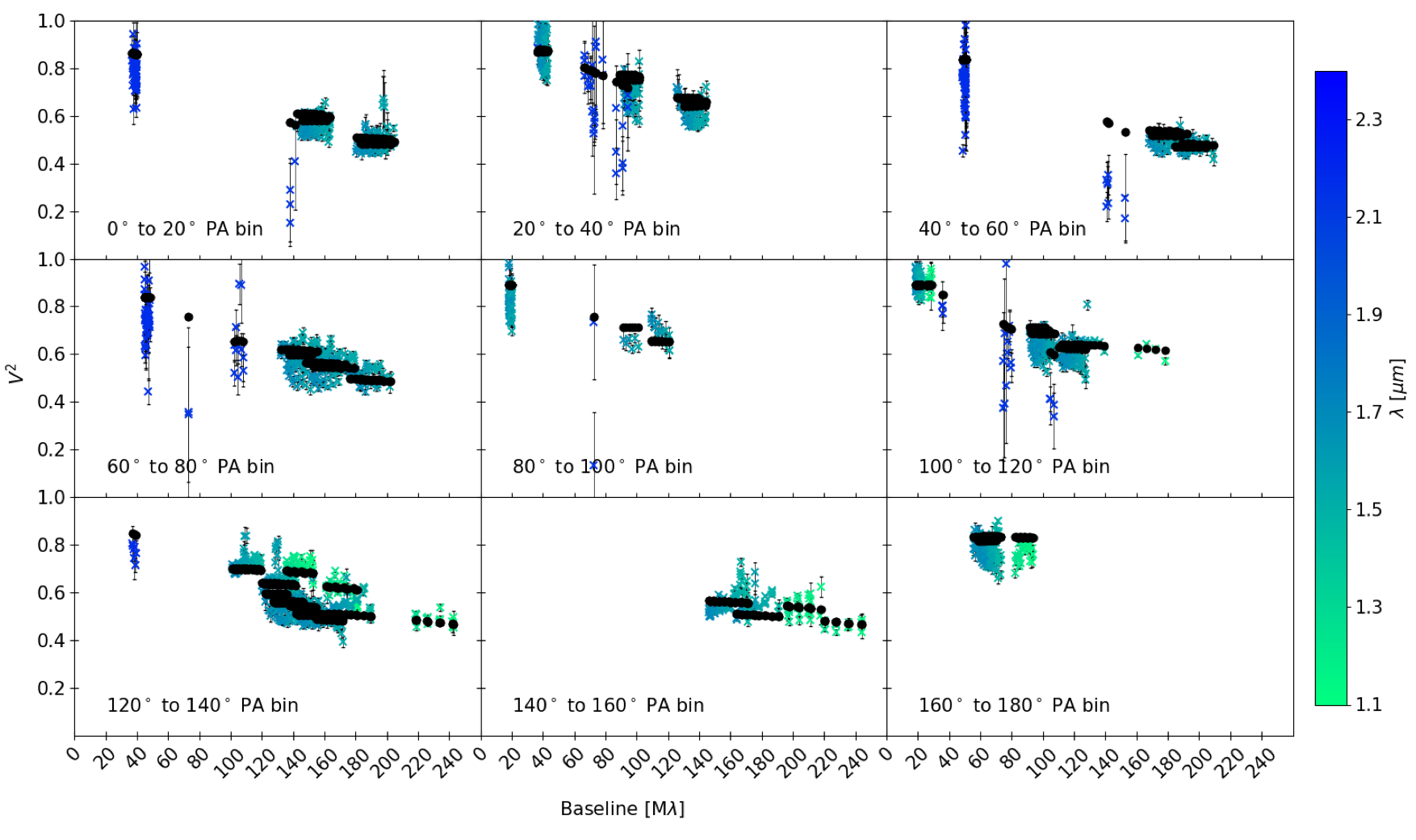}
        \caption{Squared visibilities plotted against spatial frequency, split by baseline position angle into $20^\circ$ bins. The blue/green crosses represent the interferometric observations across all instruments, coloured according to observing wavelength: dark blue is K-band data, light blue is H-band and green is J-band data. The black circles are the model visibilities of the best fit temperature gradient model corresponding to each data point.}
        \label{fig:PAmodel}
    \end{figure*}

    \renewcommand{\thefigure}{4}
    \begin{figure*}
        \centering
        \includegraphics[scale=0.6]{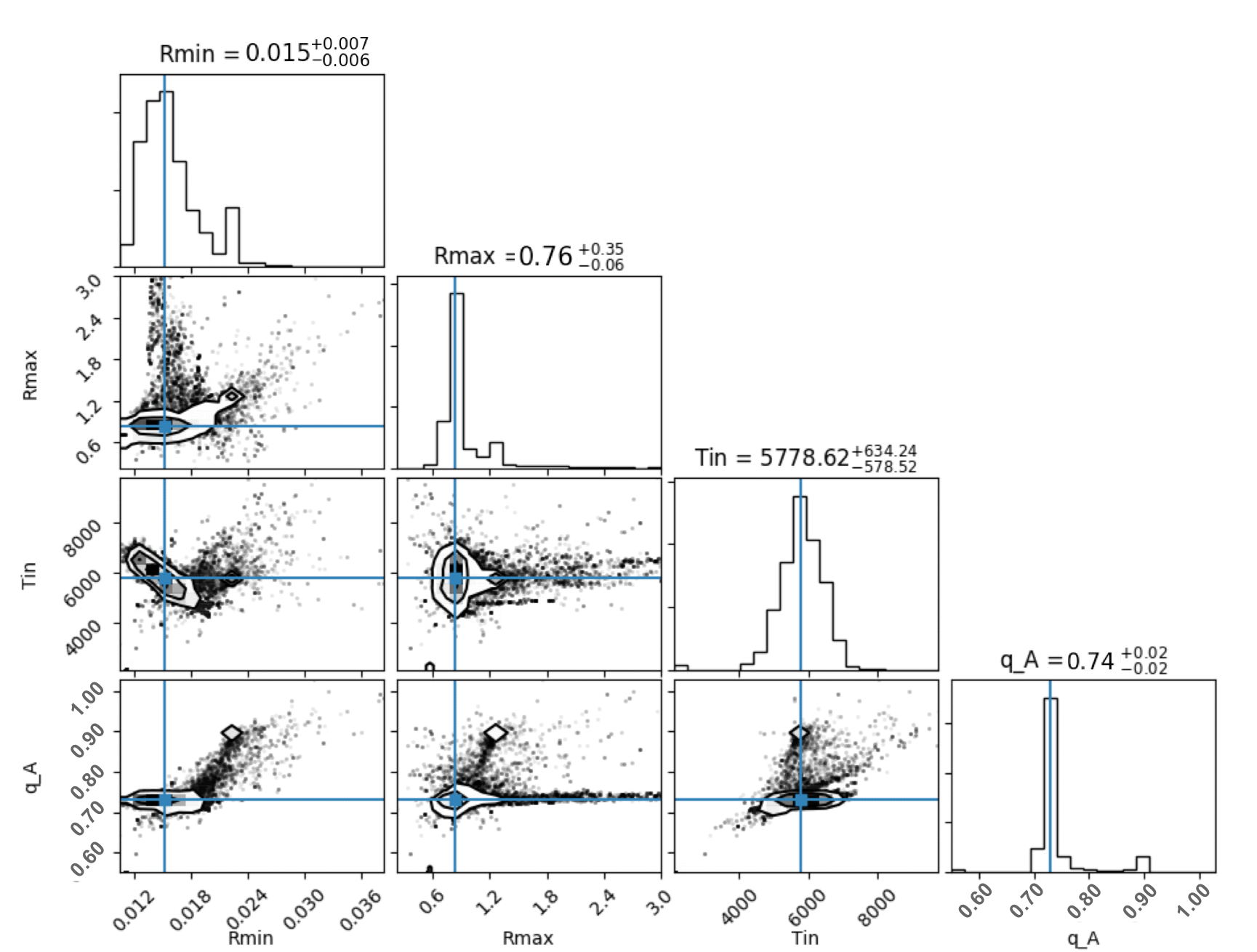}
        \caption{Posterior distributions of the parameters of the Temperature Gradient Model, produced using the corner package \citep{ForemanMackey16}. The position angle and inclination of the disk are fixed by to those obtained through Gaussian modelling. }
        \label{fig:Corner}
    \end{figure*}

    \subsection{Presentation of Results} \label{Results}
    
     The closure phases and squared visibilities obtained for all instruments are shown in Figures\,\ref{fig:CPs} and \ref{fig:PAmodel}, respectively. The visibilities are split into position angle bins of $20^\circ$ and coloured according to wavelength.
     
     The majority of our observations are contained within the MIRC-X, PIONIER and GRAVITY data which were taken over the relatively short period of 3 years. Hence the effect of photometric variability in the NIR of the object is thought to be minimal. On the other hand, the CLIMB and PTI observations were taken over a significantly longer timescale, were a small amount of photometric variability may be expected. Based on the established decreasing trend in the magnitude of 0.0125 per year in the B band, the expected drop change in the visibility based on the changing stellar-to-disk flux ratio between 1998 and 2019 is 0.045. However, we are confident that the large error bars caused by the poor signal-to-noise of these observations will successfully account for variability on the visibility and closure phase measurements. Even so, care was taken to check for time dependencies in the visibilities of baselines of similar length and position angle. K-band squared visibilities were binned according to the year of the observation and compared with each other. Each night of H-band data was compared separately to other nights. While J-band data from MIRC-X could not be compared directly, the H-band data taken simultaneously could be compared with other nights. None of these checks revealed any time variability in the data beyond the noise level, hence all interferometric data can be studied together. Also, we fitted our model to the post-2009 interferometric data alone and obtain values that are consistent on the $1.3\sigma$-level with those obtained from fitting the complete data. This confirms that any potential temporal variability does not affect our results significantly.
     
     The closure phases (Fig\,\ref{fig:CPs}) are consistent with $0^\circ$ within the error bars, indicating a centro-symmetric brightness distribution. The reduced chi-square for the closure phase measurements \citep[$\chi^2_{red-cp}$, see definition in][]{Kraus2009} for a centro-symmetric model (i.e.\ with closure phases of $0^\circ$ along all triangles) is $3.65$.
     
     \subsection{Simple Geometric Models} \label{GaussMods}
     
     \begin{table}[b!]
    \caption{\label{table:GuassFit} Best fit parameters for the Gaussian models to each of the three wavebands, independently. $\mathrm{F_{bg}}$ is the flux present in the background as a percentage of the total flux in the field of view. $\chi^2_{red}$ is the reduced chi-squared value of the best fit model for the visibilities \citep[see definition in][]{Kraus2009}. } 
    \centering
    \begin{tabular}{c c c c c c} 
        \hline
        \noalign{\smallskip}
        Band  &  FWHM & INC & PA & $\mathrm{F_{bg}}$ & $\chi^2_{red}$ \\ [0.5ex]
         & [mas] & [$^\circ$] & [$^\circ$] & [\%] & \\
        \hline
        \noalign{\smallskip}
        J & $0.38^{+0.03}_{-0.04}$ & $38.6^{+5.5}_{-9.0}$ & $37.6^{+4.0}_{-7.2}$ & $2.4^{+1.7}_{-1.2}$ & 0.22 \\
        \noalign{\smallskip}
        H & $0.41^{+0.02}_{-0.02}$ & $37.0^{+0.5}_{-0.5}$ & $41.4^{+0.7}_{-0.7}$ & $3.8^{+0.5}_{-0.5}$ & 1.68 \\
        \noalign{\smallskip}
        K & $0.60^{+0.02}_{-0.02}$ & $44.6^{+2.8}_{-3.3}$ & $32.0^{+4.2}_{-3.8}$ & $8.6^{+0.5}_{-0.4}$ & 0.76 \\
        \noalign{\smallskip}
        \hline
        \noalign{\smallskip}
    \end{tabular}
    \end{table}
     
     As a first step for interpreting the recorded interferometric observables we fitted simple geometric models to the data. This allows for the characteristic size, inclination and position angle of the object to be derived. We employ a Gaussian model within the RAPIDO (Radiative transfer and Analytic modelling Pipeline for Interferometric Disk Observations) framework. RAPIDO utilises the Markov chain Monte Carlo (MCMC) sampler {\it emcee} \citep{emcee13} to produce a fit and error estimate. The Gaussian model employed is a 'grey' model, in that it contains no spectral information. As such the three wavebands of our observations were fitted separately. In addition to a Gaussian model a secondary unresolved, extended component was also required. The free parameters of the model were the full-width-half-maximum (FWHM) of the Gaussian, the inclination (INC), minor axis position angle (PA) and the flux of the unresolved, extended component ($\mathrm{F_{bg}}$), which is measured as a percentage of the total flux in the model. 
     
     Table\,\ref{table:GuassFit} summarises the results of the Gaussian fitting. The inclination ($\sim37^\circ$) and position angle ($\sim40^\circ$) of the object are consistent across all three wavebands as expected. The size of the disk as characterised by the FWHM of the Gaussian shows an increase in size with increasing wavelength from 0.38\,mas in the J-band to 0.60\,mas in the K-band. This is expected given that longer wavelengths probes cooler regions of the disk, found at a larger distance from the central star. The flux contained within the extended component also shows a spectral dependency with a lower flux contrast at shorter wavelengths of only $2.5\%$ in the J band compared to $8.6\%$ in the K band. 
     
     A Gaussian model is intrinsically centro-symmetric and as such has a closure phase of $0^\circ$ across all triangles. This is a very good approximation to the observations, were very small closure phase signals are measured. A reduced $\chi^2_{red-cp}$ value for the closure phases of $3.65$ was calculated for a Gaussian model fitted to all data.

     \subsection{Disk Temperature Structure and Geometry} \label{GeoMod}
     
     The primary limitation of the Gaussian models employed is the lack of spectral information within the intrinsically 'grey' model. As our interferometric data covers three wavebands it is vital to account for the spectral dependency, as each wavelength channel probes a different temperature regime and hence a different disk radius.
    
     A temperature gradient model (TGM) allows for the simultaneous fitting of interferometric and photometric observables. It is built up by several rings extending from an inner radius $R_{\mathrm{in}}$ to an outer radius $R_{\mathrm{out}}$. Each ring is associated with temperature and hence flux. Therefore, a model SED can be computed by integrating over the resulting blackbody distributions for each of the concentric rings. Such a model allows us to not only to build up a picture of the temperature profile, but also approximate the position of the inner radius. The TGM is based upon $T_R = T_0(R/R_0)^{-Q}$ where $T_0$ is the temperature at the inner radius of the disk $R_0$, and $Q$ is the exponent of the temperature gradient \citep{Kreplin20,Eisner11}. A TGM represents an intrinsically geometrically thin disk. A point source is used at the centre of each model to represent an unresolved star, which is a reasonable approximation given the expected diameter of the star of $4.3\,R_\odot$ resulting in an angular diameter of $0.05\,\mathrm{mas}$ \citep{Perez20}. 
     
    \begin{table}[t!]
    \caption{\label{table:TGM_fit} Best fit parameters of the temperature gradient model. Inner disk parameters are derived in this work. Outer disk parameters ($>3\,\mathrm{au}$) are taken from \citet{Quanz06}.} 
    \centering
    \begin{tabular}{c c} 
        \hline
        \noalign{\smallskip}
        Parameter  &  Best Fit Value \\ [0.5ex]
        \hline
        \noalign{\smallskip}
        Inner Disk & $ < 3\,\mathrm{au}$\\
        \hline
        \noalign{\smallskip}
        $R_{in}$ & $0.015\pm0.007\,\mathrm{au}$ \\
        $R_{out}$ & $0.76\pm0.35\,\mathrm{au}$ \\
        $T_{in}$ & $5800\pm700\,K$ \\
        $Q$ & $0.74\pm0.02$\\
        $PA$ & $34\pm11^\circ$\\
        $INC$ & $32\pm4^\circ$\\
        \hline
        \noalign{\smallskip}
        Outer Disk & $ > 3\,\mathrm{au}$\\
        \hline
        \noalign{\smallskip}
        $T_{3\,\mathrm{au}}$ & 550\,K \\
        $Q$ & 0.53 \\
        $R_{outer}$ & 7.7\,au\\
        \hline
        \noalign{\smallskip}
    \end{tabular}
    \end{table}
    
    In the previous section we were able to reliably constrain the inclination and position angle of the disk for the first time. Our geometric modelling finds an inclination of $32\pm4^{\circ}$ and a minor axis position angle of $34\pm11^{\circ}$ measured East from North. These values were fixed in the fitting of temperature gradient model in order to reduce the number of free parameters. The fitting was undertaken using all the visibility data, shown in Figure\,\ref{fig:PAmodel} and all of the SED shown in Figure\,\ref{fig:SED}, which is constructed from the photometry detailed in Appendix\,\ref{AppB}. Interstellar extinction was taken as $E_{(B-V)} = 0.48\,(A_V = 1.4)$ as described by \citet{Pueyo12} and was used to redden the model results during fitting. The SED of FU\,Ori is unusual in the context of YSOs, in that it is almost completely dominated by disk flux even across the visible where the central star only contributes $1-2\%$. The fitting and error computation is done using the MCMC sampler corner \citep{ForemanMackey16} to produce corner plots from which parameter degeneracies and errors can be analysed. This is ideal for fitting many parameters simultaneously in a consistent manner (Fig.~\ref{fig:Corner}). 
     
    Figure\,\ref{fig:gradient} shows the best-fit temperature gradient model which corresponds to an inner disk radius of $0.015\pm0.007\,\mathrm{au}$ with a temperature of $5800\pm700\,\mathrm{K}$ and an exponent $Q = 0.74\pm\,0.02$. By design, the brightness distribution computed from our model of a geometrically thin disk with a radial temperature gradient is intrinsically centro-symmetric, meaning all closure phase measurements are equivalent to $0^\circ$. 
    
    \renewcommand{\thefigure}{5}
    \begin{figure}[b!]
        \centering
        \includegraphics[scale=0.55]{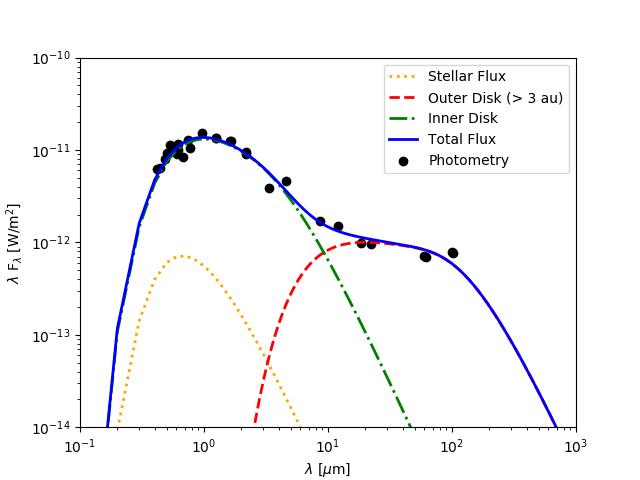}
        \caption{Spectral Energy Distribution of FU\,Ori. The black data represents the photometry points compiled in Apppendix\,\ref{AppB}. The yellow dotted line indicates the stellar flux contribution, the red dashed line the outer ($>3\,\mathrm{au}$ disk as determined by \citet{Quanz06}. The green dash-dot line represents the contribution from the inner disk, described here as a temperature gradient model. The blue line is the total flux, a sum of all components.}
        \label{fig:SED}
    \end{figure}
    
    \renewcommand{\thefigure}{6}
    \begin{figure*}[t!]
        \centering
        \includegraphics[scale=0.6]{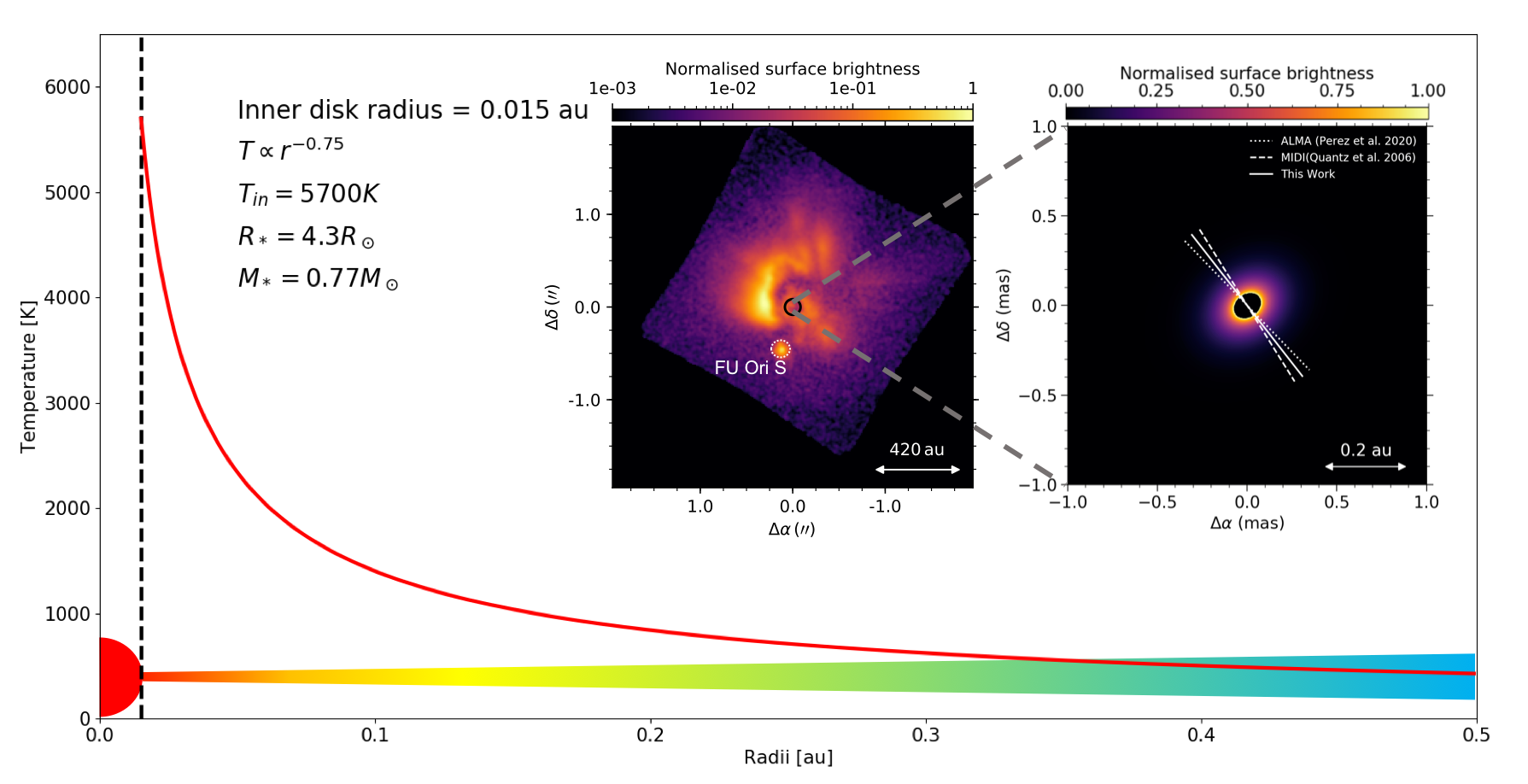}
        \caption{Temperature gradient of FU\,Ori over the inner disk. The temperature and location of the inner radius of the disk are $5800\,\mathrm{K}$ and $0.015\,\mathrm{au}$ respectively. The dashed line represents the inner disk edge. The vertical profile of the disk schematic is not to scale, nor does it represent the vertical temperature structure of the disk. {\it INSET-left:} Linearly-polarized intensity, scattered light image obtained with the Gemini Planet Imager (GPI) in the J-band \citep{Laws20}. The image is centered on FU\,Ori\,N within the black circle. FU\,Ori\,S is also visible to the South-West. {\it INSET -right:} The brightness distribution of the TGM disk model in the J-band, as derived in this work. In both panels, we plot normalised surface brightness with a logarithmic colour scale. The white lines in the image indicate the position angle of the disk minor-axis derived from other observations in literature and in this work.}
        \label{fig:gradient}
    \end{figure*}

    

    \subsection{Companion Constraints} \label{Comp}
    
    In addition to disk modelling we undertook a companion search, in order to search for a putative, previously-undetected companion within the field of view of our observations. FU\,Ori is a known binary system, however the southern component, located at a separation of $0.5"$ is outside the field of view of all interferometric observations. The companion search was undertaken using the RAPIDO model-fitting code and a companion finder extension described in \citet{Davies18}. This modul computes a grid of model, where a point source is added to the best-fit model described in section\,\ref{GeoMod}. The resulting $\chi^2_{red}$ detection significance map allows us to estimate the detection significance, or to derive an upper detection limit if no significant companion is detected.
    
    A companion search was undertaken separately on the two epochs of MIRC-X H-band data taken in November 2018 and November 2019. Only these dates were chosen as they offer good uv-coverage over the shortest possible time period (1 night), making it ideal to conduct companion searches upon. On the other hand, the supplementary data (MIRC-X J-band data, CLIMB K-band data, and PIONIER H- band data) is taken over many years and on fewer baselines, so individual epochs contain sparse uv-coverage.
    
    In order for a detection to be considered significant the p-value must be greater than $5\sigma$. This search finds that the non-zero flux solution for a companion is not significant with a p-value of $3.04$. Following the non-detection the upper limits to the flux of any companion can be calculated for each (x, y) positions in the field-of-view (grid search process described in \citet{Davies18}). The maximum flux contribution from any companion that could remain undetected by our observations is $1.3\%$ of FU Ori’s total flux in H-band within $0.5$ to $50$\,mas. Assuming a mass luminosity relation of $L\propto M^4$, this flux contrast corresponds to a maximum companion mass of $0.12\,M_\odot$.

\section{Discussion} \label{discussion}

    We have presented the first J-band interferometric observations of a young stellar object, thus demonstrating the feasibility of such observations, particularly in the context of multi-waveband interferometry. The J-band has  been a relatively untapped resource in interferometry, and the feasibility of these observations is of great interest to the wider scientific community. The J-band has the potential to not only be used in YSO studies to examine the sublimation rims and the potential for optically-thick gas inside the sublimation radius, but also in stellar photosphere studies, as a waveband which is relatively free from molecular opacities.
    
    By investigating the circumstellar environment of FU\,Ori we have explored the morphology and temperature gradient of the inner disk and greatly improved the constraints on the parameters of both the star and the disk. Geometric modelling finds a disk inclined at $32\pm4^\circ$ with a minor axis position angle of $34\pm11^\circ$. This inclination estimate is significantly more face-on than earlier estimates of $50^\circ$ and $60^\circ$ that were based on a variety of techniques, including SED analysis and near/mid-IR interferometry  \citet{Malbet05,Zhu08,Quanz06}. \citet{Calvet91} derived an inclination of $20-60^\circ$ based on the CO line-width  and \citet{Liu19} found FU\,Ori to be face-on based on NIR closure phases with GRAVITY. However, previous interferometric studies were based on very limited uv-coverage consisting solely of baselines below 100\,m and at a single wavelength, making these estimates less accurate compared to our comprehensive uv-coverage. The limitations in uv-coverage of earlier interferometric studies resulted likely also in the wide spread of minor axis position angle estimates that range from $19^\circ\pm12$ \citep{Quanz06}, $47^{\circ+7}_{-11}$ \citep{Malbet05}. In addition, many literature values are at odds with the tentative detection of a jet/outflow detected on larger scales by \citet{Takami18} and \citet{Laws20} (features C and D respectively). The putative jet/outflow feature was detected in scattered light imagery obtained with Subaru and GPI, respectively. Assuming the minor disk axis is aligned with the stellar polar axis and hence the jet position angle, a position angle of $\sim-25^\circ$ would be expected. Compared to $34\pm11^\circ$ measured here, we find a discrepancy of almost $60^\circ$. We see no evidence of such a jet in our continuum imaging on smaller scales.
    
    
    Temperature gradient models were used in order to fit the spectral dependence of the data. These models allow for the simultaneous fitting of the interferometric data and the SED. Application of these models finds a disk that extends down to the stellar surface at $0.015\pm0.007\,\mathrm{au}$, where the temperature of disk is $5800\pm700\,\mathrm{K}$. This is expected of an object that is actively accreting with such a high rate of $10^{-4}\,\mathrm{M_\odot yr^{-1}}$, and is consistent with estimates by \citet{Zhu07,Zhu08} where the disk temperature peaks at around 6000\,K. An inner disk radius equivalent to that of the star indicates boundary layer accretion directly from the disk onto the central star. Beyond the inner radius, the temperature falls off with the power-law $T\propto r^{-0.74\pm0.02}$ to an outer radius cut-off at $0.76\pm0.35\,\mathrm{au}$. The determined power-law index is consistent with the predicted temperature profile for a steady state, optically-thick accretion disk \citep{Pringle81}. A temperature gradient of this profile is only possible if viscous heating processes are present in the inner disk. Heating of flared disks by reprocessed stellar radiation alone is shown to produce temperature exponents of $q <= 0.5$ \citep{Kenyon87,Dullemond04}. Only through viscous heating can the observed temperature profile be replicated.
    
    The derived inner disk temperature gradient is in agreement with MIR work conducted with the MIDI instrument by \citet{Quanz06}, who also found a value of $Q = 0.75$ for the inner disk, although it is unclear whether their estimate was constrained mainly by interferometry or SED data. This contrasts with the outer ($>3$\,au) disk model they derive, which is also adopted here. In order to successfully fit the long wavelength SED and N-band MIDI interferometry, a temperature gradient of $T\propto r^{-0.53}$ is adopted, in good agreement to what can be found for isothermal flared disks \citep{Kenyon87}. In order to test this result, more comprehensive MIR N-band observations with the MATISSE instrument \citet{MATISSE14} at VLTI are required. 
    
    It has been proposed that FUor stars may be newborn binaries that have become bound when a small non-hierarchical multiple system breaks up \citep{Bonnell92}. In such a scenario \citet{Reipurth04} predict a close companion ($<10$\,au). Accordingly, for the FU\,Ori systems, such models would predict a third component orbiting FU\,Ori\,N in the inner 20\,mas. \citet{Malbet98} found tentative evidence for a companion located at $\sim1$\,au ($2.4$\,mas), however their data can be equally well interpreted as a circumstellar disk, which was confirmed in later work. No other studies have previously detected a companion in the inner few astronomical units around FU\,Ori\,N. We conducted a companion search based on our MIRC-X H-band visibilities and closure phases, we derive an upper limit to the flux contrast of 1.3\%, which corresponds to a maximum companion mass of 0.12\,$M_{\sun}$ in the separation range between 0.5 and 50.0 mas. Therefore, our observations do not support this scenario.

\section{Conclusions} \label{Conclusion}
    Our multi-wavelength study has probed the inner disk geometry of FU\,Ori at the highest angular resolution yet. Furthermore, we put for the first time tight constraints on the disk temperature structure in the inner astronomical unit.
    
    We summarise our conclusions as follows:
    \begin{itemize}
    
    
    \item We believe this first-of-its-kind study demonstrates a powerful and exciting new technique in the study of accretion and circumstellar disks. This allows us to directly test long-posited theoretical work with observational evidence and provide clues to the true nature of accretion and viscosity processes. 
    
    \item Temperature gradient models find an inner disk that extends down to the stellar photosphere at $0.015\pm0.007\,\mathrm{au}$ where the temperature reaches $5800\pm700\,\mathrm{K}$. This is in agreement with a heavily accreting star such as outbursting FUors and indicates boundary layer accretion processes. 
    
    \item The temperature of the inner disk falls off with a power-law $T\propto r^{-0.74\pm0.02}$. This is consistent with theoretical work for steady state, optically-thick accretion disks. Such a temperature profile is only possible if viscous heating processes are present in the inner disk.
    
    \item The inclination and position angle of the disk are tightly constrained, providing a significant improvement over literature values. An inclination of $32\pm4^\circ$ and a minor axis position angle of $34\pm11^\circ$ are found from geometric modelling.
    
    \item The minor axis position angle is around $60^\circ$ mis-aligned with the detection of a jet/outflow detected in scattered light images, assuming that the jet is perpendicular to the disk.
    
    \item No significant companion is detected within the field of view of $0.5$ to $50$\,mas. We place an upper limit of $1.3\%$ of the total $H$-band flux on an potential companion in this separation range.

    \end{itemize}
    
    Our studies demonstrates the potential of combined J and H-band interferometry to constrain the temperature structure on milliarcsecond scales. This technique enables exciting new studies on a broad range of science applications, from characterising the disks around young stellar objects to stellar surface imaging.

\begin{acknowledgements}
    We acknowledge support from an STFC studentship (No.\ 630008203) and an European Research Council Starting Grant (Grant Agreement No.\ 639889). JDM acknowledges funding from NASA NNX09AB87G, NSF NSF-ATI 1506540, and NASA XRP Grant NNX16AD43G.
    This research has made use of the VizieR catalogue access tool, CDS, Strasbourg, France. The original description of the VizieR service was published in \cite{Vizier}. 
    This work is based upon observations obtained with the Georgia State University Center for High Angular Resolution Astronomy Array at Mount Wilson Observatory.  The CHARA Array is supported by the National Science Foundation under Grant No. AST-1636624 and AST-1715788.  Institutional support has been provided from the GSU College of Arts and Sciences and the GSU Office of the Vice President for Research and Economic Development.
    MIRC-X received funding from the European Research Council (ERC) under the European Union's Horizon 2020 research and innovation programme (Grant No.\ 639889).
    Based on observations collected at the European Organisation for Astronomical Research in the Southern Hemisphere under ESO programmes 0100.C-0278(J) and 098.C-0765(C)
    
\end{acknowledgements}

\bibliographystyle{aa}
\bibliography{REF}

\begin{thebibliography}{71}
\expandafter\ifx\csname natexlab\endcsname\relax\def\natexlab#1{#1}\fi

\bibitem[{{Abrahamyan} {et~al.}(2015){Abrahamyan}, {Mickaelian}, \&
  {Knyazyan}}]{Abrahamyan15}
{Abrahamyan}, H.~V., {Mickaelian}, A.~M., \& {Knyazyan}, A.~V. 2015, Astronomy
  and Computing, 10, 99

\bibitem[{{Anugu} {et~al.}(2020){Anugu}, {Le Bouquin}, {Monnier}, {Kraus},
  {Setterholm}, {Labdon}, {Davies}, {Lanthermann}, {Gardner}, {Ennis},
  {Johnson}, {ten Brummelaar}, {Schaefer}, \& {Sturmann}}]{Anugu20}
{Anugu}, N., {Le Bouquin}, J.-B., {Monnier}, J.~D., {et~al.} 2020, arXiv
  e-prints, arXiv:2007.12320

\bibitem[{{Audard} {et~al.}(2014){Audard}, {{\'A}brah{\'a}m}, {Dunham},
  {Green}, {Grosso}, {Hamaguchi}, {Kastner}, {K{\'o}sp{\'a}l}, {Lodato},
  {Romanova}, {Skinner}, {Vorobyov}, \& {Zhu}}]{Audard14}
{Audard}, M., {{\'A}brah{\'a}m}, P., {Dunham}, M.~M., {et~al.} 2014, in
  Protostars and Planets VI, ed. H.~{Beuther}, R.~S. {Klessen}, C.~P.
  {Dullemond}, \& T.~{Henning}, 387

\bibitem[{{Bailer-Jones} {et~al.}(2018){Bailer-Jones}, {Rybizki}, {Fouesneau},
  {Mantelet}, \& {Andrae}}]{BailerJones18}
{Bailer-Jones}, C.~A.~L., {Rybizki}, J., {Fouesneau}, M., {Mantelet}, G., \&
  {Andrae}, R. 2018, \aj, 156, 58

\bibitem[{{Balbus} \& {Hawley}(1998)}]{Balbus98}
{Balbus}, S.~A. \& {Hawley}, J.~F. 1998, Reviews of Modern Physics, 70, 1

\bibitem[{{Basden} \& {Haniff}(2004)}]{Basden04}
{Basden}, A.~G. \& {Haniff}, C.~A. 2004, \mnras, 347, 1187

\bibitem[{{Beck} \& {Aspin}(2012)}]{Beck12}
{Beck}, T.~L. \& {Aspin}, C. 2012, \aj, 143, 55

\bibitem[{{Bell} \& {Lin}(1994)}]{Bell94}
{Bell}, K.~R. \& {Lin}, D.~N.~C. 1994, \apj, 427, 987

\bibitem[{{Berger} {et~al.}(2003){Berger}, {ten Brummelaar}, {Bagnuolo}, \&
  {McAlister}}]{Berger03}
{Berger}, D.~H., {ten Brummelaar}, T.~A., {Bagnuolo}, William~G., J., \&
  {McAlister}, H.~A. 2003, in Society of Photo-Optical Instrumentation
  Engineers (SPIE) Conference Series, Vol. 4838, \procspie, 974--982

\bibitem[{{Bonneau} {et~al.}(2006){Bonneau}, {Clausse}, {Delfosse}, {Mourard},
  {Cetre}, {Chelli}, {Cruzal{\`e}bes}, {Duvert}, \& {Zins}}]{Bonneau06}
{Bonneau}, D., {Clausse}, J.-M., {Delfosse}, X., {et~al.} 2006, \aap, 456, 789

\bibitem[{{Bonneau} {et~al.}(2011){Bonneau}, {Delfosse}, {Mourard}, {Lafrasse},
  {Mella}, {Cetre}, {Clausse}, \& {Zins}}]{Bonneau11}
{Bonneau}, D., {Delfosse}, X., {Mourard}, D., {et~al.} 2011, \aap, 535, A53

\bibitem[{{Bonnell} \& {Bastien}(1992)}]{Bonnell92}
{Bonnell}, I. \& {Bastien}, P. 1992, \apjl, 401, L31

\bibitem[{{Bourg{\'e}s} {et~al.}(2014){Bourg{\'e}s}, {Lafrasse}, {Mella},
  {Chesneau}, {Bouquin}, {Duvert}, {Chelli}, \& {Delfosse}}]{Bourges14}
{Bourg{\'e}s}, L., {Lafrasse}, S., {Mella}, G., {et~al.} 2014, in Astronomical
  Data Analysis Software and Systems XXIII. Proceedings of a meeting held 29
  September - 3 October 2013 at Waikoloa Beach Marriott, Hawaii, USA. Edited by
  N. Manset and P. Forshay ASP conference series, vol. 485, 2014, p.223, Vol.
  485, 223

\bibitem[{{Bouvier} {et~al.}(2007){Bouvier}, {Alencar}, {Harries},
  {Johns-Krull}, \& {Romanova}}]{Bouvier07}
{Bouvier}, J., {Alencar}, S.~H.~P., {Harries}, T.~J., {Johns-Krull}, C.~M., \&
  {Romanova}, M.~M. 2007, in Protostars and Planets V, ed. B.~{Reipurth},
  D.~{Jewitt}, \& K.~{Keil}, 479

\bibitem[{{Calvet} {et~al.}(1991){Calvet}, {Hartmann}, \& {Kenyon}}]{Calvet91}
{Calvet}, N., {Hartmann}, L., \& {Kenyon}, S.~J. 1991, \apj, 383, 752

\bibitem[{{Chambers} {et~al.}(2016){Chambers}, {Magnier}, {Metcalfe},
  {Flewelling}, {Huber}, {Waters}, {Denneau}, {Draper}, {Farrow}, {Finkbeiner},
  {Holmberg}, {Koppenhoefer}, {Price}, {Rest}, {Saglia}, {Schlafly}, {Smartt},
  {Sweeney}, {Wainscoat}, {Burgett}, {Chastel}, {Grav}, {Heasley}, {Hodapp},
  {Jedicke}, {Kaiser}, {Kudritzki}, {Luppino}, {Lupton}, {Monet}, {Morgan},
  {Onaka}, {Shiao}, {Stubbs}, {Tonry}, {White}, {Ba{\~n}ados}, {Bell},
  {Bender}, {Bernard}, {Boegner}, {Boffi}, {Botticella}, {Calamida},
  {Casertano}, {Chen}, {Chen}, {Cole}, {Deacon}, {Frenk}, {Fitzsimmons},
  {Gezari}, {Gibbs}, {Goessl}, {Goggia}, {Gourgue}, {Goldman}, {Grant},
  {Grebel}, {Hambly}, {Hasinger}, {Heavens}, {Heckman}, {Henderson}, {Henning},
  {Holman}, {Hopp}, {Ip}, {Isani}, {Jackson}, {Keyes}, {Koekemoer}, {Kotak},
  {Le}, {Liska}, {Long}, {Lucey}, {Liu}, {Martin}, {Masci}, {McLean}, {Mindel},
  {Misra}, {Morganson}, {Murphy}, {Obaika}, {Narayan}, {Nieto-Santisteban},
  {Norberg}, {Peacock}, {Pier}, {Postman}, {Primak}, {Rae}, {Rai}, {Riess},
  {Riffeser}, {Rix}, {R{\"o}ser}, {Russel}, {Rutz}, {Schilbach}, {Schultz},
  {Scolnic}, {Strolger}, {Szalay}, {Seitz}, {Small}, {Smith}, {Soderblom},
  {Taylor}, {Thomson}, {Taylor}, {Thakar}, {Thiel}, {Thilker}, {Unger},
  {Urata}, {Valenti}, {Wagner}, {Walder}, {Walter}, {Watters}, {Werner},
  {Wood-Vasey}, \& {Wyse}}]{Chambers16}
{Chambers}, K.~C., {Magnier}, E.~A., {Metcalfe}, N., {et~al.} 2016, arXiv
  e-prints, arXiv:1612.05560

\bibitem[{{Clarke} {et~al.}(2005){Clarke}, {Lodato}, {Melnikov}, \&
  {Ibrahimov}}]{Clarke05}
{Clarke}, C., {Lodato}, G., {Melnikov}, S.~Y., \& {Ibrahimov}, M.~A. 2005,
  \mnras, 361, 942

\bibitem[{{Colavita} {et~al.}(1999){Colavita}, {Wallace}, {Hines}, {Gursel},
  {Malbet}, {Palmer}, {Pan}, {Shao}, {Yu}, {Boden}, {Dumont}, {Gubler},
  {Koresko}, {Kulkarni}, {Lane}, {Mobley}, \& {van Belle}}]{Colavita99}
{Colavita}, M.~M., {Wallace}, J.~K., {Hines}, B.~E., {et~al.} 1999, \apj, 510,
  505

\bibitem[{{Cutri} {et~al.}(2014){Cutri}, {Skrutskie}, {van Dyk}, {Beichman},
  {Carpenter}, {Chester}, {Cambresy}, {Evans}, {Fowler}, {Gizis}, {Howard},
  {Huchra}, {Jarrett}, {Kopan}, {Kirkpatrick}, {Light}, {Marsh}, {McCallon},
  {Schneider}, {Stiening}, {Sykes}, {Weinberg}, {Wheaton}, {Wheelock}, \&
  {Zacarias}}]{Cutri14}
{Cutri}, R.~M., {Skrutskie}, M.~F., {van Dyk}, S., {et~al.} 2014, VizieR Online
  Data Catalog, II/328

\bibitem[{{Cutri} {et~al.}(2003){Cutri}, {Skrutskie}, {van Dyk}, {Beichman},
  {Carpenter}, {Chester}, {Cambresy}, {Evans}, {Fowler}, {Gizis}, {Howard},
  {Huchra}, {Jarrett}, {Kopan}, {Kirkpatrick}, {Light}, {Marsh}, {McCallon},
  {Schneider}, {Stiening}, {Sykes}, {Weinberg}, {Wheaton}, {Wheelock}, \&
  {Zacarias}}]{Cutri03}
{Cutri}, R.~M., {Skrutskie}, M.~F., {van Dyk}, S., {et~al.} 2003, VizieR Online
  Data Catalog, 2246

\bibitem[{{D'Alessio} {et~al.}(2005){D'Alessio}, {Calvet}, \&
  {Woolum}}]{DAlessio05}
{D'Alessio}, P., {Calvet}, N., \& {Woolum}, D.~S. 2005, in Astronomical Society
  of the Pacific Conference Series, Vol. 341, Chondrites and the Protoplanetary
  Disk, ed. A.~N. {Krot}, E.~R.~D. {Scott}, \& B.~{Reipurth}, 353

\bibitem[{{Davies} {et~al.}(2018){Davies}, {Kraus}, {Harries}, {Kreplin},
  {Monnier}, {Labdon}, {Kloppenborg}, {Acreman}, {Baron}, {Millan-Gabet},
  {Sturmann}, {Sturmann}, \& {Ten Brummelaar}}]{Davies18}
{Davies}, C.~L., {Kraus}, S., {Harries}, T.~J., {et~al.} 2018, \apj, 866, 23

\bibitem[{{Dullemond} \& {Dominik}(2004)}]{Dullemond04}
{Dullemond}, C.~P. \& {Dominik}, C. 2004, \aap, 417, 159

\bibitem[{{Eisner} \& {Hillenbrand}(2011)}]{Eisner11}
{Eisner}, J.~A. \& {Hillenbrand}, L.~A. 2011, \apj, 738, 9

\bibitem[{{ESA}(1997)}]{HIP97}
{ESA}, ed. 1997, ESA Special Publication, Vol. 1200, {The HIPPARCOS and TYCHO
  catalogues. Astrometric and photometric star catalogues derived from the ESA
  HIPPARCOS Space Astrometry Mission}

\bibitem[{{Evans} {et~al.}(2018){Evans}, {Riello}, {De Angeli}, {Carrasco},
  {Montegriffo}, {Fabricius}, {Jordi}, {Palaversa}, {Diener}, {Busso},
  {Cacciari}, {van Leeuwen}, {Burgess}, {Davidson}, {Harrison}, {Hodgkin},
  {Pancino}, {Richards}, {Altavilla}, {Balaguer-N{\'u}{\~n}ez}, {Barstow},
  {Bellazzini}, {Brown}, {Castellani}, {Cocozza}, {De Luise}, {Delgado},
  {Ducourant}, {Galleti}, {Gilmore}, {Giuffrida}, {Holl}, {Kewley}, {Koposov},
  {Marinoni}, {Marrese}, {Osborne}, {Piersimoni}, {Portell}, {Pulone},
  {Ragaini}, {Sanna}, {Terrett}, {Walton}, {Wevers}, \&
  {Wyrzykowski}}]{GAIA2Phot}
{Evans}, D.~W., {Riello}, M., {De Angeli}, F., {et~al.} 2018, \aap, 616, A4

\bibitem[{{Evans} {et~al.}(2009){Evans}, {Dunham}, {J{\o}rgensen}, {Enoch},
  {Mer{\'\i}n}, {van Dishoeck}, {Alcal{\'a}}, {Myers}, {Stapelfeldt}, {Huard},
  {Allen}, {Harvey}, {van Kempen}, {Blake}, {Koerner}, {Mundy}, {Padgett}, \&
  {Sargent}}]{Evans09}
{Evans}, Neal~J., I., {Dunham}, M.~M., {J{\o}rgensen}, J.~K., {et~al.} 2009,
  \apjs, 181, 321

\bibitem[{{Foreman-Mackey}(2016)}]{ForemanMackey16}
{Foreman-Mackey}, D. 2016, The Journal of Open Source Software, 1, 24

\bibitem[{{Foreman-Mackey} {et~al.}(2013){Foreman-Mackey}, {Hogg}, {Lang}, \&
  {Goodman}}]{emcee13}
{Foreman-Mackey}, D., {Hogg}, D.~W., {Lang}, D., \& {Goodman}, J. 2013, \pasp,
  125, 306

\bibitem[{{Gezari} {et~al.}(1993){Gezari}, {Schmitz}, {Pitts}, \&
  {Mead}}]{Gezari93}
{Gezari}, D.~Y., {Schmitz}, M., {Pitts}, P.~S., \& {Mead}, J.~M. 1993, {Catalog
  of infrared observations}

\bibitem[{{Hartmann} \& {Kenyon}(1985)}]{Hartmann85}
{Hartmann}, L. \& {Kenyon}, S.~J. 1985, \apj, 299, 462

\bibitem[{{Hartmann} \& {Kenyon}(1996)}]{Hartmann96}
{Hartmann}, L. \& {Kenyon}, S.~J. 1996, \araa, 34, 207

\bibitem[{{Henden} {et~al.}(2015){Henden}, {Levine}, {Terrell}, \&
  {Welch}}]{Henden15}
{Henden}, A.~A., {Levine}, S., {Terrell}, D., \& {Welch}, D.~L. 2015, in
  American Astronomical Society, AAS Meeting 225, id.336.16, 336.16

\bibitem[{{Herbig}(2007)}]{Herbig07}
{Herbig}, G.~H. 2007, \aj, 133, 2679

\bibitem[{{Kenyon} \& {Hartmann}(1995)}]{Kenyon95}
{Kenyon}, S. \& {Hartmann}, L. 1995, \apj, 101, 117

\bibitem[{{Kenyon} \& {Hartmann}(1987)}]{Kenyon87}
{Kenyon}, S.~J. \& {Hartmann}, L. 1987, \apj, 323, 714

\bibitem[{{Kraus} {et~al.}(2016){Kraus}, {Caratti o Garatti}, {Garcia-Lopez},
  {Kreplin}, {Aarnio}, {Monnier}, {Naylor}, \& {Weigelt}}]{kra16}
{Kraus}, S., {Caratti o Garatti}, A., {Garcia-Lopez}, R., {et~al.} 2016,
  \mnras, 462, L61

\bibitem[{{Kraus} {et~al.}(2009){Kraus}, {Hofmann}, {Malbet}, {Meilland},
  {Natta}, {Schertl}, {Stee}, \& {Weigelt}}]{Kraus2009}
{Kraus}, S., {Hofmann}, K.~H., {Malbet}, F., {et~al.} 2009, \aap, 508, 787

\bibitem[{{Kraus} {et~al.}(2020){Kraus}, {Kreplin}, {Young}, {Bate}, {Monnier},
  {Harries}, {Avenhaus}, {Kluska}, {Laws}, {Rich}, {Willson}, {Aarnio},
  {Adams}, {Andrews}, {Anugu}, {Bae}, {ten Brummelaar}, {Calvet}, {Cur{\'e}},
  {Davies}, {Ennis}, {Espaillat}, {Gardner}, {Hartmann}, {Hinkley}, {Labdon},
  {Lanthermann}, {LeBouquin}, {Schaefer}, {Setterholm}, {Wilner}, \&
  {Zhu}}]{Kraus20}
{Kraus}, S., {Kreplin}, A., {Young}, A.~K., {et~al.} 2020, Science, 369, 1233

\bibitem[{{Kraus} {et~al.}(2018){Kraus}, {Monnier}, {Anugu}, {Le Bouquin},
  {Davies}, {Ennis}, {Labdon}, {Lanthermann}, {Setterholm}, \& {ten
  Brummelaar}}]{Kraus18}
{Kraus}, S., {Monnier}, J.~D., {Anugu}, N., {et~al.} 2018, in Society of
  Photo-Optical Instrumentation Engineers (SPIE) Conference Series, Vol. 10701,
  Optical and Infrared Interferometry and Imaging VI, 1070123

\bibitem[{{Kreplin} {et~al.}(2020){Kreplin}, {Kraus}, {Tambovtseva}, {Grinin},
  \& {Hone}}]{Kreplin20}
{Kreplin}, A., {Kraus}, S., {Tambovtseva}, L., {Grinin}, V., \& {Hone}, E.
  2020, \mnras, 492, 566

\bibitem[{{Laws} {et~al.}(2020){Laws}, {Harries}, {Setterholm}, {Monnier},
  {Rich}, {Aarnio}, {Adams}, {Andrews}, {Bae}, {Calvet}, {Espaillat},
  {Hartmann}, {Hinkley}, {Isella}, {Kraus}, {Wilner}, \& {Zhu}}]{Laws20}
{Laws}, A. S.~E., {Harries}, T.~J., {Setterholm}, B.~R., {et~al.} 2020, \apj,
  888, 7

\bibitem[{{Le Bouquin} {et~al.}(2011{\natexlab{a}}){Le Bouquin}, {Berger},
  {Lazareff}, {Zins}, {Haguenauer}, {Jocou}, {Kern}, {Millan-Gabet}, {Traub},
  {Absil}, {Augereau}, {Benisty}, {Blind}, {Bonfils}, {Bourget}, {Delboulbe},
  {Feautrier}, {Germain}, {Gitton}, {Gillier}, {Kiekebusch}, {Kluska},
  {Knudstrup}, {Labeye}, {Lizon}, {Monin}, {Magnard}, {Malbet}, {Maurel},
  {M{\'e}nard}, {Micallef}, {Michaud}, {Montagnier}, {Morel}, {Moulin},
  {Perraut}, {Popovic}, {Rabou}, {Rochat}, {Rojas}, {Roussel}, {Roux},
  {Stadler}, {Stefl}, {Tatulli}, \& {Ventura}}]{PIONIER11}
{Le Bouquin}, J.~B., {Berger}, J.~P., {Lazareff}, B., {et~al.}
  2011{\natexlab{a}}, \aap, 535, A67

\bibitem[{{Le Bouquin} {et~al.}(2011{\natexlab{b}}){Le Bouquin}, {Berger},
  {Lazareff}, {Zins}, {Haguenauer}, {Jocou}, {Kern}, {Millan-Gabet}, {Traub},
  {Absil}, {Augereau}, {Benisty}, {Blind}, {Bonfils}, {Bourget}, {Delboulbe},
  {Feautrier}, {Germain}, {Gitton}, {Gillier}, {Kiekebusch}, {Kluska},
  {Knudstrup}, {Labeye}, {Lizon}, {Monin}, {Magnard}, {Malbet}, {Maurel},
  {M{\'e}nard}, {Micallef}, {Michaud}, {Montagnier}, {Morel}, {Moulin},
  {Perraut}, {Popovic}, {Rabou}, {Rochat}, {Rojas}, {Roussel}, {Roux},
  {Stadler}, {Stefl}, {Tatulli}, \& {Ventura}}]{JBLB2011}
{Le Bouquin}, J.~B., {Berger}, J.~P., {Lazareff}, B., {et~al.}
  2011{\natexlab{b}}, \aap, 535, A67

\bibitem[{{Lin} \& {Papaloizou}(1996)}]{Lin96}
{Lin}, D.~N.~C. \& {Papaloizou}, J.~C.~B. 1996, \araa, 34, 703

\bibitem[{{Liu} {et~al.}(2019){Liu}, {M{\'e}rand}, {Green}, {P{\'e}rez},
  {Hales}, {Yang}, {Dunham}, {Hasegawa}, {Henning}, {Galv{\'a}n-Madrid},
  {K{\'o}sp{\'a}l}, {Takami}, {Vorobyov}, \& {Zhu}}]{Liu19}
{Liu}, H.~B., {M{\'e}rand}, A., {Green}, J.~D., {et~al.} 2019, \apj, 884, 97

\bibitem[{{Lopez} {et~al.}(2014){Lopez}, {Lagarde}, {Jaffe}, {Petrov},
  {Sch{\"o}ller}, {Antonelli}, {Beckmann}, {Berio}, {Bettonvil}, {Glindemann},
  {Gonzalez}, {Graser}, {Hofmann}, {Millour}, {Robbe-Dubois}, {Venema}, {Wolf},
  {Henning}, {Lanz}, {Weigelt}, {Agocs}, {Bailet}, {Bresson}, {Bristow},
  {Dugu{\'e}}, {Heininger}, {Kroes}, {Laun}, {Lehmitz}, {Neumann}, {Augereau},
  {Avila}, {Behrend}, {van Belle}, {Berger}, {van Boekel}, {Bonhomme},
  {Bourget}, {Brast}, {Clausse}, {Connot}, {Conzelmann}, {Cruzal{\`e}bes},
  {Csepany}, {Danchi}, {Delbo}, {Delplancke}, {Dominik}, {van Duin}, {Elswijk},
  {Fantei}, {Finger}, {Gabasch}, {Gay}, {Girard}, {Girault}, {Gitton},
  {Glazenborg}, {Gont{\'e}}, {Guitton}, {Guniat}, {De Haan}, {Haguenauer},
  {Hanenburg}, {Hogerheijde}, {ter Horst}, {Hron}, {Hugues}, {Hummel},
  {Idserda}, {Ives}, {Jakob}, {Jasko}, {Jolley}, {Kiraly}, {K{\"o}hler},
  {Kragt}, {Kroener}, {Kuindersma}, {Labadie}, {Leinert}, {Le Poole}, {Lizon},
  {Lucuix}, {Marcotto}, {Martinache}, {Martinot-Lagarde}, {Mathar}, {Matter},
  {Mauclert}, {Mehrgan}, {Meilland}, {Meisenheimer}, {Meisner}, {Mellein},
  {Menardi}, {Menut}, {Merand}, {Morel}, {Mosoni}, {Navarro}, {Nussbaum},
  {Ottogalli}, {Palsa}, {Panduro}, {Pantin}, {Parra}, {Percheron}, {Duc},
  {Pott}, {Pozna}, {Przygodda}, {Rabbia}, {Richichi}, {Rigal}, {Roelfsema},
  {Rupprecht}, {Schertl}, {Schmidt}, {Schuhler}, {Schuil}, {Spang},
  {Stegmeier}, {Thiam}, {Tromp}, {Vakili}, {Vannier}, {Wagner}, \&
  {Woillez}}]{MATISSE14}
{Lopez}, B., {Lagarde}, S., {Jaffe}, W., {et~al.} 2014, The Messenger, 157, 5

\bibitem[{{Malbet} {et~al.}(1998){Malbet}, {Berger}, {Colavita}, {Koresko},
  {Beichman}, {Boden}, {Kulkarni}, {Lane}, {Mobley}, {Pan}, {Shao}, {Van
  Belle}, \& {Wallace}}]{Malbet98}
{Malbet}, F., {Berger}, J.~P., {Colavita}, M.~M., {et~al.} 1998, \apjl, 507,
  L149

\bibitem[{{Malbet} {et~al.}(2005){Malbet}, {Lachaume}, {Berger}, {Colavita},
  {di Folco}, {Eisner}, {Lane}, {Millan-Gabet}, {S{\'e}gransan}, \&
  {Traub}}]{Malbet05}
{Malbet}, F., {Lachaume}, R., {Berger}, J.~P., {et~al.} 2005, \aap, 437, 627

\bibitem[{{McDonald} {et~al.}(2017){McDonald}, {Zijlstra}, \&
  {Watson}}]{McDonald17}
{McDonald}, I., {Zijlstra}, A.~A., \& {Watson}, R.~A. 2017, \mnras, 471, 770

\bibitem[{{Monnier} {et~al.}(2004){Monnier}, {Berger}, {Millan-Gabet}, \& {ten
  Brummelaar}}]{Monnier04}
{Monnier}, J.~D., {Berger}, J.-P., {Millan-Gabet}, R., \& {ten Brummelaar},
  T.~A. 2004, in \procspie, Vol. 5491, New Frontiers in Stellar Interferometry,
  ed. W.~A. {Traub}, 1370

\bibitem[{{Natta} {et~al.}(2001){Natta}, {Prusti}, {Neri}, {Wooden}, {Grinin},
  \& {Mannings}}]{Natta01}
{Natta}, A., {Prusti}, T., {Neri}, R., {et~al.} 2001, \aap, 371, 186

\bibitem[{{Ochsenbein} {et~al.}(2000){Ochsenbein}, {Bauer}, \&
  {Marcout}}]{Vizier}
{Ochsenbein}, F., {Bauer}, P., \& {Marcout}, J. 2000, \aaps, 143, 23

\bibitem[{{Page} {et~al.}(2012){Page}, {Brindle}, {Talavera}, {Still}, {Rosen},
  {Yershov}, {Ziaeepour}, {Mason}, {Cropper}, {Breeveld}, {Loiseau}, {Mignani},
  {Smith}, \& {Murdin}}]{Page12}
{Page}, M.~J., {Brindle}, C., {Talavera}, A., {et~al.} 2012, \mnras, 426, 903

\bibitem[{{P{\'e}rez} {et~al.}(2020){P{\'e}rez}, {Hales}, {Liu}, {Zhu},
  {Casassus}, {Williams}, {Zurlo}, {Cuello}, {Cieza}, \& {Principe}}]{Perez20}
{P{\'e}rez}, S., {Hales}, A., {Liu}, H.~B., {et~al.} 2020, \apj, 889, 59

\bibitem[{{Pringle}(1981)}]{Pringle81}
{Pringle}, J.~E. 1981, \araa, 19, 137

\bibitem[{{Pringle} \& {Rees}(1972)}]{Pringle72}
{Pringle}, J.~E. \& {Rees}, M.~J. 1972, \aap, 21, 1

\bibitem[{{Pueyo} {et~al.}(2012){Pueyo}, {Hillenbrand}, {Vasisht},
  {Oppenheimer}, {Monnier}, {Hinkley}, {Crepp}, {Roberts}, {Brenner},
  {Zimmerman}, {Parry}, {Beichman}, {Dekany}, {Shao}, {Burruss}, {Cady},
  {Roberts}, \& {Soummer}}]{Pueyo12}
{Pueyo}, L., {Hillenbrand}, L., {Vasisht}, G., {et~al.} 2012, \apj, 757, 57

\bibitem[{{Quanz} {et~al.}(2006){Quanz}, {Henning}, {Bouwman}, {Ratzka}, \&
  {Leinert}}]{Quanz06}
{Quanz}, S.~P., {Henning}, T., {Bouwman}, J., {Ratzka}, T., \& {Leinert}, C.
  2006, \apj, 648, 472

\bibitem[{{Reipurth} \& {Aspin}(2004)}]{Reipurth04}
{Reipurth}, B. \& {Aspin}, C. 2004, \apjl, 608, L65

\bibitem[{{Saunders} {et~al.}(2000){Saunders}, {Sutherland}, {Maddox},
  {Keeble}, {Oliver}, {Rowan-Robinson}, {McMahon}, {Efstathiou}, {Tadros},
  {White}, {Frenk}, {Carrami{\~n}ana}, \& {Hawkins}}]{Saunders00}
{Saunders}, W., {Sutherland}, W.~J., {Maddox}, S.~J., {et~al.} 2000, \mnras,
  317, 55

\bibitem[{{Shakura} \& {Sunyaev}(1973)}]{Shakura73}
{Shakura}, N.~I. \& {Sunyaev}, R.~A. 1973, in X- and Gamma-Ray Astronomy,
  Vol.~55, 155

\bibitem[{{Shibazaki} \& {H{\={o}}shi}(1975)}]{Shibazaki75}
{Shibazaki}, N. \& {H{\={o}}shi}, R. 1975, Progress of Theoretical Physics, 54,
  706

\bibitem[{{Takami} {et~al.}(2018){Takami}, {Fu}, {Liu}, {Karr}, {Hashimoto},
  {Kudo}, {Vorobyov}, {K{\'o}sp{\'a}l}, {Scicluna}, {Dong}, {Tamura}, {Pyo},
  {Fukagawa}, {Tsuribe}, {Dunham}, {Henning}, \& {de Leon}}]{Takami18}
{Takami}, M., {Fu}, G., {Liu}, H.~B., {et~al.} 2018, \apj, 864, 20

\bibitem[{{Takami} {et~al.}(2014){Takami}, {Hasegawa}, {Muto}, {Gu}, {Dong},
  {Karr}, {Hashimoto}, {Kusakabe}, {Chapillon}, {Tang}, {Itoh}, {Carson},
  {Follette}, {Mayama}, {Sitko}, {Janson}, {Grady}, {Kudo}, {Akiyama}, {Kwon},
  {Takahashi}, {Suenaga}, {Abe}, {Brandner}, {Brandt}, {Currie}, {Egner},
  {Feldt}, {Guyon}, {Hayano}, {Hayashi}, {Hayashi}, {Henning}, {Hodapp},
  {Honda}, {Ishii}, {Iye}, {Kandori}, {Knapp}, {Kuzuhara}, {McElwain},
  {Matsuo}, {Miyama}, {Morino}, {Moro-Martin}, {Nishimura}, {Pyo}, {Serabyn},
  {Suto}, {Suzuki}, {Takato}, {Terada}, {Thalmann}, {Tomono}, {Turner},
  {Wisniewski}, {Watanabe}, {Yamada}, {Takami}, {Usuda}, \&
  {Tamura}}]{Takami14}
{Takami}, M., {Hasegawa}, Y., {Muto}, T., {et~al.} 2014, \apj, 795, 71

\bibitem[{{ten Brummelaar} {et~al.}(2005){ten Brummelaar}, {McAlister},
  {Ridgway}, {Bagnuolo}, {Turner}, {Sturmann}, {Sturmann}, {Berger}, {Ogden},
  {Cadman}, {Hartkopf}, {Hopper}, \& {Shure}}]{Brummelaar05}
{ten Brummelaar}, T.~A., {McAlister}, H.~A., {Ridgway}, S.~T., {et~al.} 2005,
  \apj, 628, 453

\bibitem[{{ten Brummelaar} {et~al.}(2013){ten Brummelaar}, {Sturmann},
  {Ridgway}, {Sturmann}, {Turner}, {McAlister}, {Farrington}, {Beckmann},
  {Weigelt}, \& {Shure}}]{Brummelaar13}
{ten Brummelaar}, T.~A., {Sturmann}, J., {Ridgway}, S.~T., {et~al.} 2013,
  Journal of Astronomical Instrumentation, 2, 1340004

\bibitem[{{Vorobyov} \& {Basu}(2005)}]{Vorobyov05}
{Vorobyov}, E.~I. \& {Basu}, S. 2005, \apjl, 633, L137

\bibitem[{{Vorobyov} \& {Basu}(2006)}]{Vorobyov06}
{Vorobyov}, E.~I. \& {Basu}, S. 2006, \apj, 650, 956

\bibitem[{{Zhu} {et~al.}(2007){Zhu}, {Hartmann}, {Calvet}, {Hernand ez},
  {Muzerolle}, \& {Tannirkulam}}]{Zhu07}
{Zhu}, Z., {Hartmann}, L., {Calvet}, N., {et~al.} 2007, \apj, 669, 483

\bibitem[{{Zhu} {et~al.}(2008){Zhu}, {Hartmann}, {Calvet}, {Hernand ez},
  {Tannirkulam}, \& {D'Alessio}}]{Zhu08}
{Zhu}, Z., {Hartmann}, L., {Calvet}, N., {et~al.} 2008, \apj, 684, 1281

\end{thebibliography}

\appendix
\section{List of Calibrators}\label{AppA}

\begin{table}[h]
    \caption{\label{table:Cals} Calibrators used in the reduction of interferometric data across all instruments. Details of the data reduction are described in Section\,\ref{Observations}, individual observations are shown in Table\,\ref{table:LOG} }
    
    \centering
    \begin{tabular}{c c c} 
        \hline
        \noalign{\smallskip}
        Cal Number & Calibrator ID & UDD [mas] \\ [0.5ex]
        \hline
        \noalign{\smallskip}
        (1) & \object{HD\,64515} & $0.47\pm0.01$  \\
        (2) & \object{HD\,246454} & $0.60\pm0.01$  \\
        (3) & \object{HD\,38164} & $0.43\pm0.01$  \\
        (4) & \object{HD\,28855} & $0.30\pm0.01$  \\
        (5) & \object{HD\,37320} & $0.19\pm0.01$ \\
        (6) & \object{HD\,39985} & $0.19\pm0.01$  \\
        (7) & \object{HD\,36814} & $0 .75\pm0.06$\\
        (8) & \object{HD\,38494} & $0.71\pm0.06$ \\
        (9) & \object{HD\,42618} &$0.38\pm0.01$  \\
        (10) & \object{HD\,28527} & $0.43\pm0.03$ \\
        (11) & \object{HD\,42807} &$0.45\pm0.01$  \\
        (12) & \object{HD\,35296} &$0.62\pm0.06$  \\
        (13) & \object{HD\,37147} &$0.62\pm0.01$  \\
        (14) & \object{HD\,28910} &$0.54\pm0.05$  \\
        (15) & \object{HD\,32301}& $0.50\pm0.06$ \\
        (16) & \object{HD\,50635} &$0.50\pm0.06$ \\
        (17) & \object{HD\,32923} &$0.99\pm0.11$  \\
        (18) & \object{HD\,46709} &$1.57\pm0.17$  \\
        (19) & \object{HD\,43042} &$0.68\pm0.08$ \\
        (20) & \object{HD\,35956}&$0.36\pm0.01$ \\
        \hline
    \end{tabular}
    \tablefoot{
    All uniform disk (UD) diameters quoted are in the H band and obtained from \citet{Bourges14}.
    }
    \end{table}

\section{Photometry used in the SED fitting.}\label{AppB}

\begin{table}[h]
    \caption{\label{table:Phot} Table of photometric data points used in the construction of the spectral energy distribution of FU\,Ori. Data is plotted, along with the best fitting model in Figure\,\ref{fig:SED}. Care was taken to use synchronous data wherever possible to reduce the effect of photospheric variability on the SED, the method and fitting procedure are described in Section\,\ref{GeoMod}.}
    
    \centering
    \begin{tabular}{c c c} 
        \hline
        \noalign{\smallskip}
        Wavelength [$\mathrm{\mu m}]$ & Flux [Jy] & Reference \\ [0.5ex]
        \hline
        \noalign{\smallskip}
        0.42	&	0.13	&	\citet{HIP97}\\
        0.44	&	0.02	&	\citet{Saunders00}	\\
        0.48	&	0.25	&	\citet{Henden15}	\\
        0.50	&	0.35	&	\citet{GAIA2Phot}	\\
        0.53	&	0.57	&	\citet{HIP97}	\\
        0.54	&	0.00	&	\citet{Page12}	\\
        0.55	&	0.64	&	\citet{HIP97}	\\
        0.60	&	0.55	&	\citet{McDonald17}	\\
        0.61	&	0.75	&	\citet{Chambers16}	\\
        0.62	&	0.68	&	\citet{GAIA2Phot}	\\
        0.67	&	0.70	&	\citet{GAIA2Phot}	\\
        0.74	&	1.36	&	\citet{Chambers16}	\\
        0.77	&	1.20	&	\citet{GAIA2Phot}	\\
        0.96	&	2.78	&	\citet{Chambers16}	\\
        1.24	&	3.89	&	\citet{Cutri14}	\\
        1.25	&	3.97	&	\citet{Cutri03}	\\
        1.63	&	5.46	&	\citet{Cutri03}	\\
        1.65	&	5.52	&	\citet{Cutri14}	\\
        2.16	&	5.83	&	\citet{Cutri14}	\\
        2.19    &	5.64	&	\citet{Cutri03}	\\
        3.35	&	3.91	&	\citet{Cutri14}	\\
        4.60	&	5.82	&	\citet{Cutri14}	\\
        8.61	&	4.77	&	\citet{Abrahamyan15}	\\
        11.65	&	4.16	&	\citet{Cutri14}	\\
        11.59	&	5.95	&	\citet{Gezari93}	\\
        12.00	&	5.95	&	\citet{Saunders00}	\\
        18.39	&	6.02	&	\citet{Abrahamyan15}	\\
        22.09	&	6.95	&	\citet{Cutri14}	\\
        23.88	&	14.09	&	\citet{Gezari93}	\\
        25.00	&	14.09	&	\citet{Saunders00}	\\
        60.00	&	14.29	&	\citet{Saunders00}	\\
        61.85	&	14.29	&	\citet{Gezari93}	\\
        100.00	&	26.18	&	\citet{Saunders00}	\\
        101.95	&	26.18	&	\citet{Gezari93}	\\
        \hline
    \end{tabular}
    \end{table}

\end{document}